\documentclass[journal]{IEEEtran}

\usepackage{diagbox}
\usepackage{amsmath,amssymb}
\usepackage[mathscr]{euscript}
\usepackage{amsthm}
\usepackage{boldline}

\usepackage{threeparttable}

\usepackage{color}

\usepackage{array}

\usepackage{hyperref}

\newtheorem{theorem}{Theorem}
\newtheorem{definition}{Definition}

\newtheorem{lemma}{Lemma}
\newtheorem{remarks}{Remarks}

\newcommand{\de}{\overset{\text{def}}{=}}

\begin{document}
%
\title{Second Order Asymptotics \\for Communication under Strong Asynchronism}
%
%
%

\author{\IEEEauthorblockN{Longguang Li and Aslan Tchamkerten} \\
\IEEEauthorblockA{Telecom ParisTech}
}

\maketitle

\begin{abstract}
The capacity under strong asynchronism was recently shown to be essentially unaffected by the imposed output sampling rate $\rho$ and  decoding delay $d$---the elapsed time between when information is available at the transmitter and when it is decoded. This paper examines this result in the finite blocklength regime and shows that, by contrast with capacity, the second order term in the rate expansion is sensitive to both parameters.  When the receiver must exactly locate the sent codeword, that is $d=n$ where $n$ denotes blocklength, the second order term in the rate expansion is of order $\Theta(1/\rho)$ for any $\rho=O(1/\sqrt{n})$---and $\rho =\omega(1/n)$ for otherwise reliable communication is impossible. However, if $\rho=\omega(1/\sqrt{n})$ then the second order term is the same as under full sampling and is given by a standard $O(\sqrt{n})$ term whose dispersion constant only depends on the level of asynchronism. This second order term also corresponds to the case of the slightly relaxed delay constraint $d\leq n(1+o(1))$ for any $\rho=\omega(1/n)$.
\end{abstract}


%
\IEEEpeerreviewmaketitle

\section{Introduction}
In communication under strong asynchronism \cite{tchamkerten2009communication} the information of a bursty source must be transmitted across a given discrete memoryless channel. The situation is modeled as a length $n$ codeword sent at a random time within a time window of size $e^{\alpha n}$ where $\alpha \geq 0$ quantifies the level of asynchronism in the system. Capacity as a function of the level of asynchronism is given by \cite[Corollary]{chandar2013asynchronous}
\begin{align}\label{capacity}
C(\alpha)\overset{\text{def}}{=}\max_{P: D(PW(\cdot)||W(\cdot|\star))\geq \alpha} I(P,W)
\end{align}
where $P$ and $W$ denote input distribution and the channel, respectively, where $I(P,W)$ denotes the corresponding input-output mutual information, and where $D(PW(\cdot)||W(\cdot|\star))$ denotes the relative entropy between the output distribution $\sum_x P(x)W(\cdot|x)$ and the noise distribution $W(\cdot|\star)$ corresponding to idle transmitter.
 
 Capacity turns out to mildly depend on the imposed typical detection delay $d_n$ between the instant $\nu$ when information is available at the transmitter and the instant $\tau$ when it is decoded. As long as  $$d_n\leq e^{o(n)}$$ capacity is the same as if the decoder were required to locate the codeword with minimum delay $$d_n=n,$$ that is to stop exactly at the end of the codeword transmission $\nu+n-1$ (see \cite[Corollary]{chandar2013asynchronous} and \cite[Theorem $2$]{polyanskiy2013asynchronous}).

In \cite{tchamkerten2014energy,chandar2015asynchronous} the above setting is generalized to include an output sampling constraint that imposes the receiver to observe only a fraction $\rho \leq 1$ of the channel outputs until it makes a decision.\footnote{The motivation for introducing such a constraint is to model power consumption at the receiver which, in practical systems, scales roughly linearly in the sampling rate \cite{sundstrom2009power}.} Surprisingly, it turns out that even a vanishing sampling rate $$\rho_{n}=\omega(1/n)$$ suffices to achieve the full sampling capacity $C(\alpha)$ with an almost minimal delay of $n(1+o(1))$. Conversely, if $$\rho_{n}=o(1/n)$$ the error probability is always bounded away from zero for any linear delay scheme (see \cite[Theorem $3$]{chandar2015asynchronous}).

Given the asymptotic nature of the above results a natural question is whether they are artefacts of asymptotic analysis. A first answer was provided by Polyanskiy (see \cite[Corollary $9$]{polyanskiy2013asynchronous}) for the full sampling case ($\rho_n=1$) when the synchronous capacity achieving output distribution differs from pure noise. In this case the optimization constraint in \eqref{capacity} is non-binding for any $\alpha\in [0,\bar{\alpha}) $ where $$\bar{\alpha}\overset{\text{def}}{=}D(\bar{P}W(\cdot)||W(\cdot|\star))$$ denotes the relative entropy between the (unique) capacity achieving output distribution of the synchronous channel and  $W(\cdot|\star)$---here $\bar{P}$ denotes any capacity achieving input distribution of the synchronous channel. Capacity $C(\alpha)$ is then equal to the synchronous capacity $$C=\max_P I(P,W)$$ for any $\alpha\in [0,\bar{\alpha}]$ and Polyanskiy showed that, under full sampling, even if we impose the minimal delay constraint $d_n=n$ dispersion remains equal to the dispersion of the synchronous channel. Hence for any $\alpha\in [0,\bar{\alpha}]$ asynchronism does not impact communication up to second order in the rate expansion.

In this paper we generalize the above result and quantify, in the finite blocklength regime, the impact of asynchronism, delay, and output sampling rate on communication rate. 

The results refer to two seemingly close scenarios: small delay and minimum delay. In the small delay scenario the decoder is required to achieve a delay $d_n\leq n(1+o(1))$ and in the minimum delay scenario the decoder is required to locate the codeword exactly, that is $d_n=n$. We briefly review our results:
\begin{enumerate}
\item {\it{Capacity, minimum sampling, minimum delay:}} Theorem~\ref{Them2} is a strenghtening of \cite[Theorem 3]{chandar2015asynchronous} and states that subsampling the channel outputs does not impact capacity even if the decoder is required to exactly locate the sent codeword (as opposed to achieve small delay as in \cite[Theorem 3]{chandar2015asynchronous}) whenever the sampling rate satisfies $\rho_n = \omega(1/n)$. 

\item {\it{Finite length, full sampling, minimum delay:}} Theorem~\ref{Them3} generalizes  \cite[Corollary $9$]{polyanskiy2013asynchronous} to any $\alpha\geq 0$ and shows that, under full sampling and minimum delay constraint, the second term in the rate expansion is a standard $O(\sqrt{n})$ term whose dispersion constant only depends on the level of asynchronism (and the error probability).
\item {\it{Finite length, sparse sampling, minimum delay:}} Theorem~\ref{Them4} is our first main result and says that under minimum delay constraint the second order term in the rate expansion undergoes a phase transition. If $\rho_n = \omega(1/\sqrt{n})$ the second order term remains unchanged with respect to full sampling and is of order $O(\sqrt{n})$. However, if $\rho_n = O(1/\sqrt{n})$ the second order term is of order $\Theta(1/\rho_{n})$. 
\item {\it{Finite length, sparse sampling, small delay:}} Theorem~\ref{Them5} is our second main result and says that if the decoder is allowed a small delay, then the above phase transition does not occur: for any sampling rate that is $\omega(1/n)$ the second order term is the same as under full sampling.
\end{enumerate}

\subsection*{Related works}
The present model of asynchronous communication was introduced in \cite{tchamkerten2006information}. In  \cite{chandar2008optimal, tchamkerten2009communication} the largest level of asynchronism for which reliable communication is possible was characterized. Bounds on capacity with respect to the expected detection delay were obtained in  \cite{tchamkerten2009communication} and refined in  \cite{tchamkerten2013asynchronous} which also established  suboptimality of pilot-based schemes at high asynchronism levels. Error exponents tradeoffs for false-alarms, miss-detection, and wrong decoding events were investigated in \cite{wang2011error,weinberger2014codeword} (see also \cite{weinberger2017channel}).

In \cite{chandar2013asynchronous} capacity per unit cost and, as a corollary, capacity with respect to blocklength, were characterized. In \cite{polyanskiy2013asynchronous} it was shown that requiring the decoder to locate the codeword exactly (as opposed to achieve small delay as in \cite[Corollary]{chandar2013asynchronous}) does  not affect capacity. 

In \cite{tchamkerten2014energy} the framework of \cite{chandar2013asynchronous} is extended to include an output  sampling constraint. It was shown that constraining the receiver to observe only a given fraction $\rho\in (0,1]$ of the channel outputs does not impact capacity and detection delay.  A stronger result in \cite{chandar2015asynchronous} (see Theorem $3$ therein) characterized the minimum sampling rate under which it is possible to achieve the full sampling capacity under a small delay constraint.

Beyond point-to-point communication, communication under strong asynchronism was investigated in \cite{chandar2013asynchronous} (see Remark~$2$ therein) and \cite{shahi2016capacity}  in a random access configuration and in \cite{shomorony2012bounds} in a diamond network configuration.

The above results are all asymptotic. The finite blocklength regime was first investigated in \cite{polyanskiy2013asynchronous} where it is shown that up to asynchronism level $\bar{\alpha}$ neither capacity nor dispersion are affected by asynchronism---for a finite blocklength analysis of synchronous channels we refer to \cite{polyanskiy2010channelphd}.

\subsection*{Paper organization}

 We end this section with a few notational conventions.  In Section \ref{model} we recall the sampling constrained asychronous communication setup developped in \cite{tchamkerten2014energy, chandar2015asynchronous} and review related results. In Section~\ref{section3} we present our results and in Section~\ref{proofs} we provide their proofs.

\subsection{Notation}
We use $\mathbb{P}(\cdot)$ to denote the probability of its argument and we use $\mathcal{P}^{\mathcal{X}}$, $\mathcal{P}^{\mathcal{Y}}$, and $\mathcal{P}^{\mathcal{X,Y}}$ to denote the set of distributions over finite alphabets $\mathcal{X}$, $\mathcal{Y}$, and $\mathcal{X} \times \mathcal{Y}$, respectively. We use $\mathcal{P}^{\mathcal{Y}|\mathcal{X}}$ to denote the set of conditional distributions of the form $W(y|x)$ for $(x,y) \in \mathcal{X} \times \mathcal{Y}$.

Given distributions $P \in \mathcal{P}^{\mathcal{X}}$ and $W \in \mathcal{P}^{\mathcal{Y}|\mathcal{X}}$, distribution $PW \in \mathcal{P}^{\mathcal{Y}}$ is defined as $$PW(\cdot) \overset{\text{def}}{=} \sum_{x \in \mathcal{X}}{P(x)W(\cdot|x)}.$$ 

The information divergence $D(P||Q)$ between distributions $P \in \mathcal{X}$ and $Q \in \mathcal{X}$ is denoted by $$D(P||Q) \overset{\text{def}}{=}  \sum_{x \in \mathcal{X}}{P(x)\ln{\frac{P(x)}{Q(x)}}},$$ 
and the mutual information $I(P,W)$ induced by the joint distribution $P(\cdot)W(\cdot|\cdot) \in \mathcal{P}^{\mathcal{X,Y}} $ is denoted by  
$$I(P,W) \overset{\text{def}}{=}  \sum_{x \in \mathcal{X}}{P(x)}\sum_{y \in \mathcal{Y}}{W(y|x)}  {\ln{\frac{W(y|x)}{PW(y)}}}.$$

The divergence variance between distributions  $P \in \mathcal{X}$ and $Q \in \mathcal{X}$ is denoted by (see \cite{polyanskiy2010channel})
$$V(P||Q) \overset{\text{def}}{=}  \sum_{x \in \mathcal{X}}{P(x)\left(\ln{\frac{P(x)}{Q(x)}}-D(P||Q)\right)^2}$$ 
and the conditional information variance induced by the joint distribution $P(\cdot)W(\cdot|\cdot) \in \mathcal{P}^{\mathcal{X,Y}} $ is denoted by $$V(P,W) \overset{\text{def}}{=} \sum_{x \in \mathcal{X}}{P(x)V\left(W(\cdot|x)||PW(\cdot)\right)}.$$ 
Finally, we use of the standard ``Big O'' notation for which we refer to
\url{https://en.wikipedia.org/wiki/Big_O_notation}.

\section{Model}\label{model}
We review here the asynchronous channel model with output sampling constraint developed in \cite{tchamkerten2014energy,chandar2015asynchronous}.  An asynchronous channel denoted by $$(\mathcal{X}, \star  ,W(\cdot|\cdot),\mathcal{Y})$$ consists of finite input and output alphabets $\mathcal{X}$ and $\mathcal{Y}$, respectively, and a transition probability matrix $W(\cdot|\cdot)$. Symbol $\star\in \mathcal{X}$ denotes the idle symbol. 
Without loss of generality we assume that for all $y \in \mathcal{Y}$ there is some $x \in \mathcal{X}$ such that $W(y|x) > 0$. 

A codebook $\mathcal{C}$ consists of $M\geq 1$ length $n$ codewords composed of symbols from $\mathcal{X}$. Communication rate is defined as

$$R \overset{\text{def}}{=} \frac {\ln M}{n}.$$

A random and uniformly chosen message $m$ is made available at the transmitter at a random time $\nu$ that is uniformly distributed over $$\{1,2,\ldots,A=e^{n\alpha}\}$$ where $A$ quantifies the asynchronism level and where $\alpha$ denotes the corresponding asynchronism exponent. Random variables $m$ and $\nu$ are assumed to be independent. 

 The transmitter starts to transmit codeword $c^{n}(m) \in \mathcal{C}$ across channel $W$ at a time $\sigma(\nu,m)$ that may causally depend on both $\nu$ and $m$, {\it{i.e.}}, such that
 $$\nu \leq \sigma(\nu,m) \leq A  \, \, \, \, \text{almost surely.}$$ 
Outside the transmission interval, {\it{i.e.}}, before time $\sigma$ and after time $\sigma + n-1$, the receiver observes only pure noise, that is i.i.d. samples distributed according to $W(\cdot|\star)$. 
Specifically, conditioned on the event that codeword $c^n(m)$ starts being sent at time $\{\sigma = t \}$, $t \in \{1,2,\dots, A\}$,  output process $\{Y_{i}\}$ is distributed according to
\begin{align*}
&\mathbb{P}_{Y_{1}^{A+n-1}|X^{n},\sigma}\left(y_{1}^{A+n-1}|c^{n}(m),t\right) \\
&= \left(\prod_{i < t,i > t + n - 1}W(y_{i}|\star)\right) \prod_{t \leq i \leq t + n - 1}W(y_{i}|c_{i}(m)).
\end{align*}

Without knowing $\nu$ the decoder operates according to
\begin{itemize}
\item
 a sampling strategy,
\item
a stopping time defined on the sampled process,
\item
 a decoding function defined on the stopped sampled process.
\end{itemize}
A sampling strategy consists of ``sampling times'' which are
defined as an ordered collection of random time indices
$$\mathcal{S} = \{(S_{1},\dots,S_{l}) \subseteq \{1,\dots,A+n-1\}: S_i < S_j,i < j\}$$
where $S_j$ is interpreted as the $j$th sampling time. The sampling
strategy may or may not be adaptive. It is non-adaptive
when $\mathcal{S}$ is independent of $Y_{1}^{A+n-1}$.
The strategy is adaptive when the sampling times are functions
of past observations. This means that $S_1$ is an arbitrary value
in $\{1,\dots, A + n -1\}$, possibly random but independent of $Y_{1}^{A_{n}+n-1}$, and for $j \geq 2$,
$$S_{j}=g_{j}(\{Y_{S_{i}} \}_{i < j} )$$
for some (possibly random) function
$$g_{j}:\mathcal{Y}^{j-1} \rightarrow \{S_{j-1}+1,\dots,A+n-1\}.$$
Notice that $l$, the total number of output samples, may be
random under adaptive sampling but also under non-adaptive
sampling since the strategy may be randomized (but still
independent of the channel outputs $Y_{1}^{A+n-1}$). 

The set of sampling times up to a time $t\geq 1$ is defined as
$$\mathcal{S}^{t}\overset{\text{def}}{=}\{S_i: S_i\leq t\}$$ and the corresponding set of sampled channel outputs is defined as
$$\mathcal{O}^{t} \overset{\text{def}}{=} \{Y_{S_{i}}: S_{i} \leq t\}.$$
The receiver decodes by means of
a sequential test $(\tau, \phi_{\tau} )$ where $\tau$ denotes a stopping (decision)
time with respect to $\{\mathcal{O}^{t}\}$
and where $\phi_{\tau}$ denotes a decoding function defined on the
stopped sampled output process.\footnote{Recall that a (deterministic or randomized) stopping time $\tau$ with respect
to a sequence of random variables $	Y_1,Y_2,\dots$ is a positive, integer-valued,
random variable such that the event $\{\tau = t\}$, conditioned on the realization
of $Y_1, Y_2,\dots,Y_t$ , is independent of the realization of $Y_{t+1}, Y_{t+2},\dots$, for all
$t \geq 1$.}The decoding function $\phi_{\tau}$ maps the set of samples observed until time $\tau$ to a message estimate, that is\footnote{Notice that since there are at most $A + n - 1$ sampling times we have that both $\tau$ and
$|\mathcal{O}^{\tau}|$ are upperbounded by $A + n - 1$.}
$$\phi_{\tau}:\mathcal{Y}^{|\mathcal{O}^{\tau}|} \rightarrow \{1,2,\dots,M\}$$
$$\mathcal{O}^{\tau} \mapsto \phi_\tau(\mathcal{O}^{\tau}).$$
Hence a message $m$ is wrongly decoded if event
$$\tilde{\mathcal{E}}_{m} \overset{\text{def}}{=} \{ \phi_\tau({\mathcal{O}}^{\tau}) \neq m \}$$ 
happens.

A code $$(\mathcal{C},(\mathcal{S},\tau,\phi_{\tau}))$$ is defined as a codebook and a
decoder composed of a sampling strategy, a decision time, and
a decoding function. Throughout the paper, whenever clear
from context, we often refer to a code using the codebook
symbol $\mathcal{C}$ only, leaving out explicit references sampling
strategy, decoding time, and decoding function.


%


Next we define  error probability of a code by penalizing wrong decoding, detection delay, and sampling rate at the decoding instant. We introduce two types of error. 
\begin{definition}[Error events] Fix $d \in \mathbb{N}$ and $0 \leq \rho \leq 1$. Given $(d,\rho)$ and a code $\mathcal{C}$ we define error event 
\begin{align} \label{E7071}
{\mathcal{E}_{1,m}}\overset{\text{def}}{=}\tilde{\mathcal{E}}_{m} \cup \{ \tau - \nu + 1 > d \} \cup \{ |{\mathcal{S}}^{\tau}|/\tau > \rho \},
\end{align}
and given $\rho$ and a code $\mathcal{C}$ we define error event
\begin{align} \label{E707}
{\mathcal{E}_{2,m}}\overset{\text{def}}{=}\tilde{\mathcal{E}}_{m} \cup \{ \tau - \nu + 1 \neq n \} \cup \{ |{\mathcal{S}}^{\tau}|/\tau > \rho \}.
\end{align}

For each of the two types of error the maximum error probability of a code $\mathcal{C}$ operating at asynchronism level $A$ is defined as
\begin{align*}
 \max_{m}\mathbb{P}(\mathcal{E}_{i,m}) \overset{\text{def}}{=} \max_{m}\frac{1}{A}\sum_{t = 1}^{A}\mathbb{P}_{m,t}(\mathcal{E}_{i,m})\qquad i\in \{1,2\}.
\end{align*}
Subscript ``$m,t$'' indicate conditioning on the event that message $m$ is available at time $\nu = t$ at the transmitter.
\end{definition}

Observe that ${\mathcal{E}_{1,m}}$ penalizes decoding if delay is larger than~$d$. By contrast ${\mathcal{E}_{2,m}}$ penalizes decoding whenever delay is not exactly equal to the blocklength.  In particuar notice that 
 \begin{align}\label{e1le2}{\mathcal{E}_{1,m}}\subseteq {\mathcal{E}_{2,m}} 
\end{align}
 whenever $d\geq n$.
\begin{definition}[Achievable Rate]Fix $\alpha \geq 0$ and fix two sequences  $\{d_{n}\}$ and $\{\rho_n\}$ such that $d_{n} \in \mathbb{N}$ and such that $\{\rho_n\}$ is non-increasing and satisfies $0 \leq \rho_{n} \leq 1$. Rate $R$ is said to be achievable (with respect to error event \eqref{E7071} or \eqref{E707}) if for any $\varepsilon > 0$ and $n$ sufficiently large there exists a code $\mathcal{C}$ that
\begin{enumerate}
\item operates under asynchronism level $A = e^{n\alpha}$,
\item yields a rate at least $R-\varepsilon$,
\item achieves a maximum probability of error at most $\varepsilon$.
\end{enumerate}
\end{definition}
\begin{definition}[Asynchronous Capacity] Fix $\alpha \geq 0$ and fix two sequences  $\{d_{n}\}$ and $\{\rho_n\}$ such that $d_{n} \in \mathbb{N}$ and such that $\{\rho_n\}$ is non-increasing and satisfies $0 \leq \rho_{n} \leq 1$. 
Asynchronous capacity $C_1(\alpha,\{\rho_n\},\{d_n\})$ with respect to error event \eqref{E7071} is the supremum of the set of achievable rates at $(\alpha,\{\rho_n\},\{d_n\})$. Asynchronous capacity $C_2(\alpha,\{\rho_n\})$ with respect to error event \eqref{E707} is the supremum of the set of achievable rates at $(\alpha,\{\rho_n\})$.

\end{definition}
Because of \eqref{e1le2} we have
\begin{align}\label{c1c2}
C_1(\alpha,\{\rho_n\},\{d_n\})\geq C_2(\alpha,\{\rho_n\})
\end{align}
for any $\alpha\geq 0$, any $\{\rho_n\}$, and any $ \{d_n\}$ such that $d_n\geq n$.

\begin{remarks}\label{rema1}
Note that samples occurring after time $\tau$ play no role in our performance metrics since both events ${\mathcal{E}_{1,m}}$ and ${\mathcal{E}_{2,m}}$ are functions of $\mathcal{O}^{\tau}$. Hence, without loss of generality, for the rest of the paper we assume that the last sample is taken at time $\tau$. In particular, we have
\begin{align}\label{assumption}
|\mathcal{S}^{\tau}| \geq |\mathcal{S}^t|  \qquad 1 \leq t \leq A+n-1.
\end{align}
\end{remarks}

The following theorem characterizes capacity under full sampling. In this regime delay plays little role since capacity remains the same whether delay is minimal or subexponential in the blocklength. 
  \begin{theorem}[Full sampling, minimum delay, see Corollary \cite{chandar2013asynchronous} and Theorem $2$ \cite{polyanskiy2013asynchronous}] \label{Them1}
 Fix $\alpha \geq 0$. Then,
 \begin{itemize}
 \item  whenever $n \leq d_n \leq e^{o(n)}$ we have $$C_1(\alpha,\{\rho_n=1\},\{d_n\})=C_2(\alpha,\{\rho_n=1\})=C(\alpha)$$
 where $C(\alpha)$ is defined in \eqref{capacity};
\item to achieve  $C(\alpha)$ it is necessary that $d_n\geq n(1-o(1))$ under error event \eqref{E7071}.
 \end{itemize}
 \end{theorem}
Theorem~\ref{Them1} says that under full sampling the capacity of subexponential delay codes is $C(\alpha)$. Moreover, and somewhat surprisingly, to achieve $C(\alpha)$ it is sufficient to consider codes that locate the codeword exactly, thereby essentially achieving minimum delay by the second part of the theorem.

Comparing expression \eqref{capacity} with the synchronous capacity $$C=\max_P I(P,W)$$ we deduce that there exists a critical asynchronism exponent $\bar{\alpha}$  such that for any $\alpha \in [0, \bar{\alpha}]$ capacity is not impacted by asynchronism, that is $$C(\alpha) = C.$$ This critical exponent is given by
\begin{align}\label{albar}
\bar{\alpha} \overset{\text{def}}{=}
 D\left(\bar{P}W\left(\cdot)||W(\cdot|\star\right)\right)
 \end{align} where $\bar{P}$ is any input distribution that achieves the synchronous capacity $C$ and where $\bar{P}W\left(\cdot\right)$ denotes the corresponding (unique) capacity achieving output distribution. In particular, we have $\bar{\alpha}>0$ if and only if the pure noise distribution $W(\cdot|\star)$ differs from the capacity achieving output distribution of the synchronous channel.

Intuition suggests that constraining the output sampling rate impacts the decoder's ability to detect the sent message and ultimately contributes to reduce capacity. Remarkably, a sparse output sampling generally does not impact capacity even under a small delay constraint:
  \begin{theorem}[Minimum sampling, small delay, see Theorem $5$ \cite{chandar2015asynchronous}] \label{Them22}
 Fix $\alpha \geq 0$ and  fix $\{\rho_n\}$ such that $\rho_n=\omega(1/n)$. Then for some $d_n = n(1+o(1))$
$$C_1(\alpha,\{\rho_n\},\{d_n\}) = C(\alpha).$$
Conversely, if $\rho_n=o(1/n)$ and $d_n=\Theta(n)$ then $$C_1(\alpha,\{\rho_n\},\{d_n\})=0.$$
 \end{theorem}
Hence, there is a phase transition at sampling rate of order $\Theta(1/n)$.  If $\rho_n=o(1/n)$ communication is impossible for any linear delay code---even for a two-message code the error probability is bounded away from zero. However, if $\rho_n=\omega(1/n)$ then it is possible to achieve the full sampling capacity.

The next definitions pertain to finite length analysis.

\begin{definition}[$(n,\varepsilon,\alpha, d,\rho)$- and $(n,\varepsilon,\alpha, \rho)$-code] 
Fix $\alpha \geq 0$, $d \in \mathbb{N}$, $0 \leq \rho \leq 1$, and $0 < \varepsilon < 1$. Code ${\mathcal{C}}$ is an $(n,\varepsilon,\alpha, d,\rho)$ code if it operates at asynchronism exponent $\alpha$ and achieves error probability with respect to error event \eqref{E7071} not exceeding~$\varepsilon$.
The maximum cardinality of a codebook of a $(n,\varepsilon,\alpha, d,\rho)$ code is denoted by $M^{*}_1(n,\varepsilon,\alpha,d,\rho)$. 

Similarly, code ${\mathcal{C}}$ is an $(n,\varepsilon,\alpha,\rho)$ code if it operates at asynchronism exponent $\alpha$ and achieves  error probability with respect to error event \eqref{E707} not exceeding~$\varepsilon$. The maximum cardinality of a codebook of a $(n,\varepsilon,\alpha, \rho)$ code is denoted by $M^{*}_2(n,\varepsilon,\alpha,\rho)$. 

Whenever clear from context we shall shorten the notation to $M^{*}_1(n,\varepsilon)$ and $M^{*}_2(n,\varepsilon)$, respectively.
\end{definition}

 \begin{definition}[Dispersion]
Given $\alpha \geq 0$ and $0<\varepsilon<1$, the $\varepsilon$-channel dispersion $V_{\varepsilon}(\alpha)$ of the synchronous channel with input distribution $P$ subject to 
\begin{align}\label{condition}
D(PW(\cdot)||W(\cdot|\star)) \geq \alpha
\end{align}
  is defined as
\begin{align}\label{dispersion}
V_{\varepsilon}(\alpha) \overset{\text{def}}{=}  \begin{cases}
V_{\min}(\alpha) &\mbox{if $\varepsilon < 1/2$} \\
V_{\max}(\alpha) &\mbox{if $\varepsilon \geq 1/2$}
\end{cases}
\end{align}
where $$V_{\min}(\alpha) \overset{\text{def}}{=} \min \limits_{P \in \Pi_{\alpha}}V(P,W)$$ $$V_{\max}(\alpha) \overset{\text{def}}{=} \max \limits_{P \in \Pi_{\alpha}}V(P,W)$$
and where $$\Pi_{\alpha} \overset{\text{def}}{=}\{P: D(PW(\cdot)||W(\cdot|\star)) \geq \alpha, I(P,W) = C(\alpha)\}.$$
\end{definition}

\begin{remarks}
Throughout the paper we always assume without loss of generality that the asynchronism threshold $\alpha$ is strictly below $$\alpha(W)=\max_x D(W(\cdot|x)||W(\cdot|\star)),$$ the synchronization threshold of the channel. Recall (see \cite{chandar2008optimal, tchamkerten2009communication}) that $\alpha(W)$ quantifies the largest level of asynchronism below which reliable communication is possible under full sampling. Above $\alpha(W)$ any decoder won't locate the sent codeword within a window of size $e^{o(n)}$ with probability approaching one.
\end{remarks}

 \section{Results}\label{section3}

The first result provides a slight improvement upon Theorem~\ref{Them22} and \cite[Theorem~$2$]{polyanskiy2013asynchronous} and says that above the critical sampling rate of order $1/n$ it is possible to achieve both the full sampling  capacity and minimum delay (instead of small delay):
\begin{theorem}[Minimum sampling, minimum delay]\label{Them2}
 Fix $\alpha \geq 0$ and  fix $\{\rho_n\}$ such that $\rho_n=\omega(1/n)$. Then
$$C_2(\alpha,\{\rho_n\}) = C(\alpha).$$
\end{theorem}

The next results refer to rate expansion in the finite blocklength regime and represents a non-asymptotic version of Theorem~\ref{Them1}:
\begin{theorem}[Finite length, full sampling, minimum delay] \label{Them3}
Fix $\alpha\geq 0$ and $\varepsilon \in (0,1)$. If $\rho_n=1$ we have\footnote{We use $Q(x)$ to denote the standard $Q$-function $Q(x)=(1/\sqrt{2\pi})\int_{x}^\infty \exp(-t^2/2)dt$.}
\begin{equation} \label{E205}
\ln{M^{*}_2(n,\varepsilon)} = nC(\alpha) - \sqrt{nV_{\varepsilon}(\alpha)}Q^{-1}(\varepsilon) + O(\ln{n}).
\end{equation}
\end{theorem}

When $\alpha \in (0, \bar{\alpha}]$ ($\bar{\alpha}$ is defined in \eqref{albar}) Theorem \ref{Them3}  reduces to \cite[Corollary 9]{polyanskiy2013asynchronous} and implies that asynchronism impacts neither capacity nor dispersion. However, when $\alpha >\bar{\alpha}$ asynchronism impacts both capacity and dispersion.  

Theorem~\ref{Them4} is a non-asymptotic version of Theorem~\ref{Them2} and a sparse sampling version of Theorem~\ref{Them3}:
\begin{theorem}[Finite length, sparse sampling, minimum delay]\label{Them4}
Fix $\alpha \geq 0$ and $\varepsilon \in (0,1)$. If $\rho_n=\omega(1/\sqrt{n})$ then
\begin{equation}\label{E302}
\ln{M^{*}_2(n,\varepsilon)} = nC(\alpha) - \sqrt{nV_{\varepsilon}(\alpha)}Q^{-1}(\varepsilon)+o(\sqrt{n}). 
\end{equation}
Conversely, if $\rho_{n} = O(1/\sqrt{n})$ (and $\rho_{n} = \omega(1/n)$) then
 \begin{equation}\label{E303}
\ln{M^{*}_2(n,\varepsilon)} = nC(\alpha) - \Theta(1/\rho_{n})+O(\sqrt{n}).  
\end{equation}
\end{theorem}
Hence, for a given level of asynchronism and under the minimal delay constraint, capacity undergoes a phase transition when the sampling rate is of order $1/n$ (Theorem~\ref{Them2}) and the second order term in the rate expansion undergoes a phase transition when the sampling rate is of order $1/\sqrt{n}$. If $\rho_n=\omega(1/\sqrt{n})$ the second order term is captured by a ``classical'' $O(\sqrt{n})$ term whose dispersion constant only depends on the level of asynchronism. If instead $\rho_n=O(1/\sqrt{n})$ (and $\rho_{n} = \omega(1/n)$) the second order term is captured by the inverse of the sampling rate.

For synchronous communication it is well-known that the second order term in the rate expansion is $O(\sqrt{n})$ universally over discrete memoryless channels. So the reader may wonder why this no longer holds in the current context of asynchronous communication. 

The reason is that the receiver may miss part of the sent codeword. Indeed, when the decoder is required to achieve minimum delay it can be shown that the decoder will miss at least order $\Theta(1/\rho_n)$ symbols from the sent codeword. Therefore,  the receiver observes a truncated version of the sent codeword of size  at most $$\tilde{n}=n-\Theta(1/\rho_n).$$ This, by the rate expansion for the synchronous channel, implies that the second order term in the asynchronous rate expansion is $\Omega(1/\rho_n)$ if $\rho_n=O(1/\sqrt{n})$. Finally, through an achievability scheme it can be shown that the second order term is $\Theta(1/\rho_n)$ if $\rho_n=O(1/\sqrt{n})$.

The above argument turns out to critically depend on the minimum delay constraint. If we only slightly relax this constraint to a small delay constraint of $n(1+o(1))$ then the second order term in the rate expansion is again of order $O(\sqrt{n})$:

\begin{theorem}[Finite length, sparse sampling, small delay]\label{Them5}
Fix $\alpha \geq 0$ and $\varepsilon \in (0,1)$. If $\rho_n=\omega(1/n)$ then for some delay $d_n = n(1+o(1))$ we have 
\begin{equation}\label{E301}
\ln{M^{*}_1(n,\varepsilon)} = nC(\alpha) - \sqrt{nV_{\varepsilon}(\alpha)}Q^{-1}(\varepsilon)+O(\ln{n}).  
\end{equation}
\end{theorem}
The reason why a relaxation of the codeword location constraint allows to increase rate is that it reduces the effective level of asynchronism since the decoder needs only to locate the sent codeword within a time window of size $n(1+o(1))$ instead of $n$. This slight asynchronism reduction turns out to be sufficient to recover the $\sqrt{n}$ in the second order term of the rate expansion as we will see in the proof of Theorem~\ref{Them5}.  

It should be stressed that in Theorem \ref{Them5} the exact form of $\rho_n$ impacts the $o(1)$ term in the delay.

\begin{table}
\renewcommand{\arraystretch}{1.5}
\caption{Capacity and second order term as functions of the asynchronism exponent $\alpha$ and sampling rate $\rho_n$ for exact codeword location.}
\label{T101}
\centering
\begin{threeparttable}
\begin{tabular}{| c V{2} c|c| }
 \hline 
\diagbox{$\rho_n$}{$\alpha$} & $(0,\bar{\alpha}]$ &$(\bar{\alpha}, \alpha(W))$\\ [1ex]
\hlineB{2}
 $\omega(1/\sqrt{n})$   & $C,\sqrt{V_{\varepsilon}^{\text{sync}}n}$  &$C(\alpha),\sqrt{V_{\varepsilon}(\alpha)n}$\\ 
 \hline
 $o(1/\sqrt{n})$ and $\omega(1/n)$ &$C,\Theta(1/\sqrt{\rho_n})$  & $C(\alpha),\Theta(1/\sqrt{\rho_n}) $   \\ 
 \hline
\end{tabular}
\end{threeparttable}
\end{table}

\begin{table}
\renewcommand{\arraystretch}{1.5}
\caption{Capacity and second order term as functions of the asynchronism exponent $\alpha$ and the sampling rate $\rho_n$ for small delay constraint $n(1+o(1))$.}
\label{T102}
\centering
\begin{threeparttable}
\begin{tabular}{| c V{2} c|c| }
 \hline 
\diagbox{$\rho_n$}{$\alpha$} & $(0,\bar{\alpha}]$ &$(\bar{\alpha}, \alpha(W))$\\ [1ex]
\hlineB{2}
 $\omega(1/\sqrt{n})$   & $ C,\sqrt{V_{\varepsilon}^{\text{sync}}n}$   &$C(\alpha),\sqrt{V_{\varepsilon}(\alpha)n}$\\ 
 \hline
 $o(1/\sqrt{n})$ and $\omega(1/n)$ &$C,\sqrt{V_{\varepsilon}^{\text{sync}}n}$  & $C(\alpha),\sqrt{V_{\varepsilon}(\alpha)n}$   \\ 
 \hline
\end{tabular}
\end{threeparttable}
\end{table}

Tables  \ref{T101} and  \ref{T102} summarize the main results and show the capacity and the second order term in the rate expansion as a function of asynchronism and output sampling rate for minimum delay and small delay constraints. The constant $V_\varepsilon^{\text{sync}}$ denotes the dispersion of the synchronous channel, that is $V_\varepsilon^{\text{sync}}=V_\varepsilon(\alpha=0)$, and recall that $C$ denotes the capacity of the synchronous channel, that is $C=C(\alpha=0)$.


\section{Proofs}\label{proofs}
\subsection{Proof of Theorem \ref{Them2}}
\subsubsection{Achievability}
We show that any rate $R<C(\alpha)$ is achievable under error event ${\mathcal{E}_{2,m}}$ as long as the sampling rate satisfies $\rho_n=\omega(1/n)$. To do this 
we use a variant of the multiphase scheme introduced in \cite[Theorem 5]{chandar2015asynchronous} which allows to achieve minimum delay (instead of small delay) and which is better suited for rate expansion analysis. This variant removes the codewords' preambles and differs in both the last phase of the scheme and in the criteria under which phases change.

Fix $\alpha>0$, $A = e^{n\alpha}$, $R<C(\alpha)$, and let $P$  be any distribution in $ \Pi_{\alpha}$. Let $v_1$ denote the variance of random variable $$\ln \frac{PW(\cdot)}{W(\cdot|\star)}$$ under distribution $PW(\cdot)$ and let $$v_2 \overset{\text{def}}{=} V(P\times W || P \times PW).$$
Let $$\rho_{n}= f(n)/n$$ where $f(n) = \omega(1)$ and $f(n) \leq n$. 
 Furthermore, we define the following parameters which pertain to the coding scheme described below:
 \begin{align}
 \beta_i &= \Delta_{i}(D(PW(\cdot)||W(\cdot|\star))-\delta_1)\quad i=1,2,\ldots, \ell-1 \label{E813} \\
 \beta_{\ell} &= (n-\Delta(n))(D(PW(\cdot)||W(\cdot|\star))-\delta_1) \label{E814} \\
 \gamma &= (n-\Delta(n))(I(P,W)-\delta_2)  \label{E815}
\end{align}
where constant $\delta_1$ is arbitrary in 
$$(0,D(PW(\cdot)||W(\cdot|\star))-\alpha)$$
and where constant $\delta_2$ is arbitrary in
$$(0,I(P,W)-R).$$

Next, define the exponentially increasing sequence $\{\Delta_{i}\}_{i=1}^\ell$ recursively as
$$\Delta_{1} \overset{\text{def}}{=} f(n)^{\delta}$$ 
\begin{align}\label{E4011}
 \Delta_{i} \overset{\text{def}}{=}\min\{e^{c_{i}\Delta_{i-1}}, n \}
 \end{align}
 for $i = 2,\cdots,\ell$ where $\ell$ denotes the smallest integer such that $\Delta_{\ell} = n$, where $c_{i}$ is arbitrary in $$(0,\beta_{i}/\Delta_{i}),$$ 
 and where $\delta \in (0,1/2)$ is arbitrary.

Finally, we shall use the shorthand notation $r(Y^k)$ and $i(X^k;Y^k)$ to denote the  log-likelihood ratios
\begin{align}\label{E401}
r(Y^k) \overset{\text{def}}{=} \ln{\frac{ PW({Y}^{k}) } {W({Y}^{k}|\star^{k})}}
\end{align}
and
\begin{align}\label{E402}
i(X^k;Y^k) \overset{\text{def}}{=} \ln{ \frac{P(Y^k|X^k)}{PW(Y^k)  }}
\end{align}
where $(X^k,Y^K)$ is distributed according to the $k$-fold product distribution of $P(\cdot)W(\cdot|\cdot)$, and where
$$PW(Y^k = {y}^{k}) \overset{\text{def}}{=} \sum_{x^k \in \mathcal{X}^k}{P(x^k)W(y^k|x^k)}.$$

 Codewords for each message $m \in \{1,2\cdots,M\}$ are independently generated so that 
 $$C^{n}(m) \overset{\text{def}}{=}  C_{1}(m)C_{2}(m) \cdots C_{n}(m)$$
 is i.i.d. according to distribution $P$. The transmission start time $\sigma$ is set to be equal to $\nu$.

 The receiver parses the set of time indices $\{1,2,\cdots,A\}$ into $$ A / \Delta(n) $$ blocks of length $$\Delta(n) \overset{\text{def}}{=} n/f(n)^{1-2\delta}.$$ We denote $t_{i}$ as the starting time of the $i$th block.\footnote{We implicitly assume here that $\Delta(n)$ is an integer. Throughout the analysis we shall ignore issues related to the rounding of non-integer quantities as they play no role asymptotically.}


The decoder operates as follows. It takes samples at the beginning of each block and checks for the presence of a message using a cascade of $\ell$ binary hypothesis tests, of increasing reliability, each acting as a confirmation test of the previous test. 
If any of these tests fail, samples are skipped till the beginning of the next block. If all tests are positive, the decoder stops and outputs a message estimate. Details follow.

 At time $t_1$, {\it{i.e.}}, at the beginning of the first block, the decoder starts with the first detection phase by taking $\Delta_{1}$ samples and checks whether  $$r({Y}^{\Delta_{1}}) \geq \beta_{1}.$$ If not, the decoder skips samples till $t_2$, the beginning of the second block, and restarts the procedure by repeating the first phase. If the above inequality holds the decoder switches to a confirmation second phase and checks if $$r({Y}^{\Delta_{2}}) \geq \beta_{2}.$$ If the test is negative, samples are skipped till time $t_2$ and if the test is positive the decoder switches to a confirmation third phase. 

If $\ell-1$ consecutive phases are passed, the decoder moves to a last decoding phase. The decoder takes $\Delta_{\ell}=n$ samples and checks whether within the $$\sum_{i=1}^{\ell}{\Delta_{i}}$$ samples observed in the current block there are $n-\Delta(n)$ consecutive samples such that 
\begin{align}\label{EA101}
r({Y}^{n-\Delta(n)}) \geq \beta_{\ell}
\end{align}
and for which there is a unique codeword $C^{n}(\hat{m})$ that satisfies 
\begin{align}\label{EA102}
i(C^{n-\Delta(n)}(\hat{m});{Y}^{n-\Delta(n)}) \geq \gamma.
\end{align}
 If so the decoder stops and declares message $\hat{m}$. If more than one such codeword exists, the decoder declares one at random. If one of the above conditions is not satisfied, the decoder skips samples until the next transmission time and restarts the procedure afresh with the first phase. If no codeword is found by time $A+n-1$ the decoder declares a random message.

We now compute the error probability of the above strategy assuming message $m$ is sent. 

The error probability can be bounded as 
\begin{align}
\mathbb{P}(\mathcal{E}_{2,m}) &\leq \mathbb{P}(\mathcal{E}_{I} \displaystyle \mathop{\cup} \mathcal{E}_{II} \displaystyle \mathop{\cup} \mathcal{E}_{III} \displaystyle \mathop{\cup} \mathcal{E}_{IV} \mathop{\cup}  \mathcal{E}_{V}) \notag \\
&\leq \mathbb{P}(\mathcal{E}_I)+\mathbb{P}(\mathcal{E}_{II})+\mathbb{P}(\mathcal{E}_{III})+\mathbb{P}(\mathcal{E}_{IV})+\mathbb{P}(\mathcal{E}_{V}) \label{100}
\end{align}
where events $\mathcal{E}_{I}$, $\mathcal{E}_{II}$, $\mathcal{E}_{III}$, $\mathcal{E}_{IV}$, and $\mathcal{E}_{V}$ are defined as
\begin{align*}
 \mathcal{E}_{I}&=\{\tau < \nu\} \\
\mathcal{E}_{II}&=\{\nu\leq \tau\leq \nu+n-1, \hat{m}\ne m\}\\
\mathcal{E}_{III}&=\{\nu\leq \tau\leq \nu+n-1, |{\mathcal{S}}^{\tau}|/\tau > \rho_n\}\\
 \mathcal{E}_{IV}&=\{\nu\leq \tau\leq \nu+n-2, \hat{m}=m\}\\
 \mathcal{E}_{V}&=\{\tau > \nu + n - 1\}.
\end{align*}

For event $\mathcal{E}_{I}$ we have
%
\begin{align}
\mathbb{P}(\mathcal{E}_{I}) &= \mathbb{P}(\tau < \nu) \leq A \cdot  \mathbb{P}_\star(r({Y}^{n-\Delta(n)})\ge \beta_{\ell})\label{E405b} \\
&\leq e^{-(n-\Delta(n))(D(PW(\cdot)||W(\cdot|\star)) - \alpha - \delta_1)} \label{E405}
\end{align}
by \eqref{EA101} and a union bound over time indices, where $\mathbb{P}_\star$ denotes probability under the pure noise distribution (that is vector ${Y}^{n-\Delta(n)}$ is i.i.d. according to distribution $W(\cdot|\star)$).


By union bound over times indices and the fact that the set of codewords $\{C^{n}(\hat{m}):\hat{m}\ne m\}$ are independent of the channel outputs, we have
\begin{align}
\mathbb{P}(\mathcal{E}_{II})  &\leq  \sum_{j=1}^{n} { \sum_{\hat{m} \neq m}{ \mathbb{P}(i(C^{n-\Delta(n)}(\hat{m});{Y}_{j}^{n-\Delta(n)+j-1}) \geq \gamma )}} \notag \\
&\leq nMe^{-\gamma} \label{E408b}\\
&\leq ne^{-(n-\Delta(n))(I(P,W)-R-\delta_2)}.
 \label{E408}
\end{align}
In Appendix \ref{AppA} we show that
%
%
%
\begin{align}
\mathbb{P}(\mathcal{E}_{III})  
&\leq \frac{1}{\sqrt{A}} + \frac{ \left(1+f(n)^{-\delta} \sum_{i=2}^{\ell}e^{-(\beta_{i}-c_{i})}\right) }{f(n)^{\delta}(1-n^{2}/\sqrt{A})}\notag \\
 &\leq e^{-n\alpha/2}+O(f(n)^{-\delta}) \label{E414}
\end{align}
and in Appendix \ref{AppB} we show that
\begin{align}
\mathbb{P}(\mathcal{E}_{IV}) &\leq \Delta(n)e^{-\gamma} + (n-\Delta(n))Me^{-\gamma}\label{E708b} \\
&\leq ne^{-(n-\Delta(n))(I(P,W)-R-\delta_2)}. \label{E708}
\end{align}
To bound $\mathbb{P}(\mathcal{E}_{V})$ we bound the probability of failing any of the $\ell$ tests during information transmission as 
\begin{align}
\mathbb{P}&(\mathcal{E}_{V}) \leq \mathbb{P} (i(C^{n-\Delta(n)}(m);{Y}^{n-\Delta(n)}) < \gamma) + \notag \\
 &\sum_{i=1}^{\ell-1}{\mathbb{P}(r({Y}^{\Delta_{i}}) < \beta_{i})} + \mathbb{P}(r({Y}^{n-\Delta(n)}) < \beta_{\ell})  \label{E415b} \\
&\leq  (v_1/\delta_1^{2}+v_2/\delta_2^{2})/(n-\Delta(n)) + v_1/\delta_1^{2}\sum_{i=1}^{\ell-1}\frac{1}{\Delta_{i}} 
 \label{E415}
\end{align}
where the second inequality follows from Chebyshev's inequality.

From \eqref{100}, \eqref{E405}, \eqref{E408}, \eqref{E414}, \eqref{E708}, \eqref{E415} we have that 
$\mathbb{P}(\mathcal{E}_{2,m})$ vanishes as $n\to \infty$, thereby proving the theorem.

\subsubsection{Converse}
From Theorem~\ref{Them22}
we have that $$C_1(\alpha,\{\rho_n\},\{d_n\})=C(\alpha)$$
for any given $\rho_n=\omega(1/n)$ and for some $d_n=n(1+o(1))$. 
Therefore by \eqref{c1c2} we get 
$$C_2(\alpha,\{\rho_n\})\leq C(\alpha).$$ 
This concludes the converse.
 ~\qed
\subsection{Proof of Theorem \ref{Them3}}

\subsubsection{Achievability}
To establish achievability we use a simplified variant of the achievability scheme of Theorem~\ref{Them22} where encoding is kept unchanged and where the decoder operates only according to the last phase---hence $\ell=1$.  At each time $t$ the decoder checks whether the last $n$ consecutive channel outputs $Y^n$ are such that 
\begin{align}
r({Y}^n) \geq \beta_{1}
\end{align}
and for which there is a unique codeword $C^{n}(\hat{m})$ that satisfies 
\begin{align}
i(C^{n}(\hat{m});{Y}^{n}) \geq \gamma.
\end{align}
 If so the decoder stops and declares message $\hat{m}$. If more than one such codeword exists, the decoder declares one at random. If one of the above conditions is not satisfied, the decoder moves to time $t+1$ and repeats the procedure. If no codeword is found by time $A+n-1$ the decoder declares a random message.

For the error probability, since $\rho_n=1$, we have
\begin{align}
\mathbb{P}(\mathcal{E}_{2,m})\leq \mathbb{P}(\mathcal{E}_{I})+\mathbb{P}(\mathcal{E}_{II})+\mathbb{P}(\mathcal{E}_{IV})+ \mathbb{P}(\mathcal{E}_{V})
\end{align}
and using \eqref{E405b}, \eqref{E408b}, \eqref{E708b} (with $\Delta(n)=0$), and \eqref{E415b}
\begin{align}
\mathbb{P}(\mathcal{E}_{2,m}) \leq A \cdot  \mathbb{P}_{\star}(r(&{Y}^{n})\ge \beta_{1}) + 2nMe^{-\gamma}   \notag \\
+\mathbb{P}(r({Y}^{n}) &< \beta_{1}) +\mathbb{P} (i(C^{n}(m);{Y}^{n}) < \gamma).  \label{E901}
\end{align}
For any $0 < \varepsilon < 1$,  $P\in \Pi_{\alpha}$ and 
\begin{align}
 \beta_1 &= n(D(PW(\cdot)||W(\cdot|\star))-\delta_1) \notag \\
\log{M} &= nC(\alpha) - \frac{3}{2}\log{n} - \sqrt{nV_{\varepsilon}(\alpha)} \times \notag \\
&\qquad \qquad \qquad \times Q^{-1}(\varepsilon -\frac{B+3}{\sqrt{n}} - \frac{v_1}{n\delta_{1}^2} ) \label{E821} \\
\gamma&=\log{M}+\frac{3}{2}\log{n} \notag 
\end{align}
where $B$ denotes the Berry-Esseen constant \cite{polyanskiy2010channel},
the four terms on the left-hand side of \eqref{E901} can be upper bounded as
\begin{align}
A \cdot \mathbb{P}_{\star}(r({Y}^{n})\ge \beta_{1}) \leq e^{-n\delta_1} &\leq \frac{1}{\sqrt{n}}  \label{EA902}\\
2nMe^{-\gamma} &\leq \frac{1}{\sqrt{n}} \label{EA903} \\ 
\mathbb{P} (i(C^{n}(m);{Y}^{n}) < \gamma) &\leq \varepsilon -\frac{2}{\sqrt{n}} - \frac{v_1}{n\delta_1^2} \label{EA904} \\
\mathbb{P}(r({Y}^{n}) < \beta_{1}) &\leq \frac{v_1}{n\delta_1^2} \label{EA905} 
\end{align}
 where the second inequality in \eqref{EA902} and inequality \eqref{EA903} hold for $n$ large enough, where \eqref{EA904} follows from the Berry-Esseen Theorem, and where \eqref{EA905} follows from Chebyshev's inequality. 

Hence, for $n$ large enough
 $$\mathbb{P}(\mathcal{E}_{2,m}) \leq \varepsilon$$ 
and the proof is completed by applying the Taylor expansion in equality \eqref{E821}. This shows that
$$\ln{M^{*}_2(n,\varepsilon)} \geq  nC(\alpha) - \sqrt{nV_{\varepsilon}(\alpha)}Q^{-1}(\varepsilon) + O(\ln{n}).$$
\subsubsection{Converse} 
To prove the converse, we need the following Lemma which applies to the synchronous channel. The proof is deferred to Appendix~\ref{AppD}. 
\begin{lemma}\label{Lem2}
Fix $\alpha \geq 0$, $0 < \varepsilon <1$, and let $P$ be a distribution satisfying \eqref{condition}. The maximum cardinality $ M_{\text{sync}}^{*}(n,\varepsilon)$ of a length $n$ code, of constant composition $P$, and maximal error probability at most $\varepsilon$ over a synchronous channel $W$ satisfies\footnote{Since the channel is synchronous error probability is defined in the standard way with respect to event $\{\hat{m}\ne m\}$ only and considers neither oversampling nor delay.}
\begin{align*}
\ln{M_{\text{sync}}^{*}(n,\varepsilon)} = nC(\alpha) - \sqrt{nV_{\varepsilon}(\alpha)}Q^{-1}(\varepsilon) + O(\ln{n}).
\end{align*}
\end{lemma}

Consider an $(n, \varepsilon, \alpha, \rho=1, d=n)$ code with cardinality $M$. Through a standard expurgation argument  (see \cite[Chap.11, p.349]{cover2012elements}), there exists a constant composition subcode of cardinality $M^{\prime}$ that satisfies 
\begin{align} \label{E05231}
\ln{M^{\prime}} \geq \ln{M} - c\ln{n}
\end{align}
where $c > 0$ is a constant independent of the chosen code. When $n$ is sufficiently large, the empirical distribution of this subcode satisfies \eqref{condition} by the proof of \cite[Theorem 1, eq. (151)]{polyanskiy2013asynchronous}. Moreover, since the error probability of this subcode is at most $\varepsilon$ over the asynchronous channel it is no larger than $\varepsilon$ when used over the synchronous channel. Therefore  Lemma~\ref{Lem2} applies and we have
\begin{align}\label{E05232}
M^{\prime} \leq M_{\text{sync}}^{*}(n,\epsilon).
\end{align}
Substituting \eqref{E05232} into \eqref{E05231}, we get
$$\ln{M} \leq nC(\alpha) - \sqrt{nV_{\varepsilon}(\alpha)}Q^{-1}(\varepsilon) + O(\ln{n})$$
which concludes the converse.~\qed

\subsection{Proof of Theorem \ref{Them4}}
\subsubsection{$\rho_{n} = \omega(1/\sqrt{n})$} We show that  \eqref{E302} holds for any $\rho_{n} = \omega(1/\sqrt{n})$. Since the converse is implies by the converse of Theorem~\ref{Them3} (which holds under full sampling), we only need to establish achievability. 

The achievability scheme is identical to the achievability scheme used to establish Theorem \ref{Them2},  with the same parameters except for the following:
\begin{align}
 \beta_{\ell} &= (n-\Delta(n))(D(PW(\cdot)||W(\cdot|\star))-\delta_1) \notag  \\
\ln{M} &= (n-\Delta(n))C(\alpha) -\frac{3}{2}\ln{n} -  \notag \\
 &\sqrt{(n-\Delta(n))V_{\varepsilon}(\alpha)}\times Q^{-1}\bigg(\varepsilon -\frac{B+4}{\sqrt{n}}  -f(n)^{-\frac{\delta}{2}} -  \notag \\
 &\qquad \qquad -\frac{v_1}{\delta_1^{2}(n-\Delta(n))} - \frac{v_1}{\delta_1^{2}}\sum_{i=1}^{\ell-1}{\frac{1}{\Delta_{i}}} \bigg) \label{E416} \\
\gamma&=\ln{M}+\frac{3}{2}\ln{n} \notag
\end{align}
and where $f(n)=\omega(1)$ such that $\rho_n=f(n)/\sqrt{n}$.

Using similar calculations as in \eqref{E405}-\eqref{E415} we get
\begin{align*}
\mathbb{P}(\mathcal{E}_{I})&\leq \frac{1}{\sqrt{n}}  \\
\mathbb{P}(\mathcal{E}_{II}) &\leq \frac{1}{\sqrt{n}}\\
\mathbb{P}(\mathcal{E}_{III})&\leq f(n)^{-\delta/2} \\
\mathbb{P}(\mathcal{E}_{IV}) &\leq \frac{2}{\sqrt{n}} \\
\mathbb{P}(\mathcal{E}_{V})&\leq \varepsilon-\frac{4}{\sqrt{n}}  - 
 f(n)^{-\frac{\delta}{2}}
 \end{align*}
and therefore 
 $$\mathbb{P}(\mathcal{E}_{2,m}) \leq \varepsilon$$ 
for $n$ large enough.  Finally, a Taylor expansion in equality \eqref{E416} gives \eqref{E301}.
\subsubsection{$\rho_{n} = O(1/\sqrt{n})$} We first show that \eqref{E303} holds for any $\rho_n=o(1/\sqrt{n})$ and then handle the case when $\rho_n=\Theta(1/\sqrt{n})$. 

Suppose $\rho_{n} = 1/(\sqrt{n}\cdot f(n))$ where $f(n) = \omega(1)$ and $f(n) = o(\sqrt{n})$. For achievability, we consider the same coding scheme as in the achievability part of Theorem~\ref{Them4} with the only difference that now $\Delta(n)$ is redefined as $$\Delta(n) = \sqrt{n}f(n)g(n)$$
where $g(n)$ is arbitrary and satisfies $g(n) = \omega(1)$ and $g(n)=o(\sqrt{n}/f(n))$. Similar calculations as for Theorem~\ref{Them4} show that there exists a code of cardinality $M$ that satisfies
 \begin{align}
 \ln{M} = nC(\alpha)-\sqrt{n}f(n)g(n)C(\alpha) + O(\sqrt{n})  \label{E781}
 \end{align}
and such that the error probability is no larger than $\varepsilon$. 
Therefore
\begin{align}
\ln{M^*_2(n,\varepsilon)} \geq nC(\alpha)-\sqrt{n}f(n)g(n)C(\alpha) + O(\sqrt{n}) . \label{E7811}
\end{align}

For the converse, let $0 < \varepsilon < 1$ and  suppose code ${\mathcal{C}}$ is an $(n,\varepsilon, \alpha, \rho_n)$ code with $M$ codewords.

We have  
\begin{align}
 \varepsilon&>\mathbb{P}({\mathcal{E}}_{2,m})\notag \\
 & \geq \mathbb{P}(\tilde{\mathcal{E}}_{m}\cup \{\tau\ne \nu+n-1\}) \notag \\
 &\geq  \mathbb{P}(\tilde{\mathcal{E}}_{m}\cup \{\tau\ne \nu+n-1\},\mathcal{D}(\nu,\mathcal{S}^{\tau})) \notag \\
 &= \mathbb{P}(\mathcal{D}(\nu,\mathcal{S}^{\tau})) \cdot \mathbb{P}(\tilde{\mathcal{E}}_{m}\cup \{\tau\ne \nu+n-1\}\big\vert\mathcal{D}( \nu,\mathcal{S}^{\tau})) \notag \\
 &\geq (1-o(1)) \cdot \mathbb{P}(\tilde{\mathcal{E}}_{m}\cup \{\tau\ne \nu+n-1\}\big\vert\mathcal{D}( \nu,\mathcal{S}^{\tau})) \label{E605}
\end{align} 
where 
$$\mathcal{D}(\nu,\mathcal{S}^{\tau}) \overset{\text{def}}{=} \left\{ \mathcal{S}^{\tau} \cap \{\nu,\nu +1, \cdots,\nu +\sqrt{n}f(n)h(n)\} = \emptyset\right\} 
  $$
with $h(n)$ arbitrarily chosen that satisfies $h(n)=o(1)$ and $h(n) = \omega(1/f(n))$. The last inequality in \eqref{E605} holds by Lemma~\ref{intersect} with $q(n)=\sqrt{n}f(n)h(n)$. 

From \eqref{E605}
we have 
$$\mathbb{P}(\tilde{\mathcal{E}}_{m}\cup \{\tau\ne \nu+n-1\}\big\vert\mathcal{D}( \nu,\mathcal{S}^{\tau}))\leq \varepsilon+o(1).$$
Therefore there exists a code of length at most $n - \sqrt{n}f(n)h(n)$ and of cardinality $M$ for which the error probability can be made no larger than $\varepsilon+o(1)$ when used synchronously. Using the same argument as the one after Lemma~\ref{Lem2} (expurgation argument) we get
\begin{align}
\ln M\leq nC(\alpha) - \sqrt{n}f(n)h(n)C(\alpha) + O(\sqrt{n}) \label{E606B}
\end{align}
and therefore
\begin{align}
\ln M^*_2(n,\varepsilon)\leq nC(\alpha) - \sqrt{n}f(n)h(n)C(\alpha) + O(\sqrt{n}). \label{E606B}
\end{align}
Since $g(n)=\omega(1)$ in \eqref{E781} and $h(n)=o(1)$ in \eqref{E606B} are arbitrary and recalling that $\rho_n=1/(\sqrt{n}f(n))$ we deduce that
\begin{align}
\ln M^*_2(n,\varepsilon)= nC(\alpha) -\Theta(1/\rho_n) + O(\sqrt{n}). \label{E606BB}
\end{align}
This concludes the case when $\rho_n=o(1/\sqrt{n})$.
 
For the case $\rho_n = \Theta(1/\sqrt{n})$, to show that \eqref{E303} holds it suffices to prove that
$$\ln{M^{*}_2(n,\varepsilon)} = nC(\alpha) - \Theta(\sqrt{n})+o(\sqrt{n}).$$

The converse is immediate by Lemma~\ref{Lem2} and the standard expurgation argument.

For achievability we consider the same coding scheme as in the achievability part of Theorem~\ref{Them4} with $\Delta(n)$ redefined as $$\Delta(n) = \sqrt{n}g(n)$$
where $g(n) = \omega(1)$ and $g(n)=o(\sqrt{n})$. Similar calculations as for Theorem~\ref{Them4} show that there exists a code such that the second order of its rate expansion is $$O(\sqrt{n}g(n)).$$
Since the above argument holds for any $g(n)=\omega(1)$ this concludes the case $\rho_n=\Theta(1/\sqrt{n})$.
\subsection{Proof of Theorem \ref{Them5}}
\subsubsection{Achievability}
The achievability scheme is identical to the achievability scheme used to establish Theorem~\ref{Them2} except for the following differences.

The transmission time $\sigma$ takes values over multiples of $\Delta(n)$. If 
$$(i-1)\Delta_n<\nu <i \Delta_n$$ then we set 
$$\sigma=t_i=i \Delta_n.$$
In other words, compared to the achievability scheme of Theorem~\ref{Them2} transmission can now be delayed by up to $\Delta_n=o(n)$.

The decoding process only differs in the last phase. In the $\ell$th phase, the decoder checks whether there are  $n$ consecutive samples within the past $$\sum_{i=1}^{\ell}{\Delta_{i}}$$ samples (as opposed to the last $\Delta_{\ell}$ samples) that satisfy 
\begin{align*}
r({Y}^{n}) \geq \beta_{\ell}
\end{align*}
and for which there is a unique codeword $C^{n}(\hat{m})$ that satisfies 
\begin{align*}
i(C^{n}(\hat{m});{Y}^{n}) \geq \gamma.
\end{align*}
If so the decoder stops and declares message $\hat{m}$. If more than one such codeword exists, the decoder declares one at random. If one of the above conditions is not satisfied, the decoder skips samples until the next transmission time and restarts the procedure afresh with the first phase. If no codeword is found by time $A+n-1$ the decoder declares a random message.

Considering the achievability scheme of Theorem~\ref{Them2} with the above two modifications---transmission time and last phase---we bound the error probability as $${\mathbb{P}}(\mathcal{E}_{1,m}) \leq {\mathbb{P}}( \mathcal{E}_{I}) +{\mathbb{P}}( \mathcal{E}_{II}) +{\mathbb{P}}( \mathcal{E}_{III})+{\mathbb{P}}(  \mathcal{E}_{V})$$
where events ${\mathcal{E}_{I}}$, ${\mathcal{E}_{II}}$, ${\mathcal{E}_{III}}$, ${\mathcal{E}_{V}}$ are defined as in the proof of Theorem~\ref{Them2} (see after~\eqref{100}) but with $\nu$ changed to $\sigma$. Bounding these probabilities similarly as in the achievability part of Theorem~\ref{Them4} (with the same parameters) the proof can be concluded.
\subsubsection{Converse}
Apply the converse of Theorem \ref{Them3} which holds under full sampling and therefore also under sparse sampling. ~\qed

\section{Concluding remarks}
In this paper we provided a non-asymptotic characterization of communication rate as a function of asynchronism, delay, and output sampling rate. The main conclusion is on how asynchronism and receiver sampling rate impact capacity and the second term in the rate expansion. Asynchronism impacts capacity and the dispersion constant for both the minimum and the small delay constraints. However, asynchronism does not impact the order of the second term (which remains $\sqrt{n}$). By contrast, sampling rate never impacts capacity but can, depending on the delay constraint, significantly impact the order of the second term in the rate expansion.

\appendices

\section{Proof of Inequality \eqref{E414}}\label{AppA}
This proof follows the similar arguments as in the achievability part of \cite[Theorem 5]{{chandar2015quickly}}.
For event $\mathcal{E}_{III}$, we have
\begin{align}
\mathbb{P}(\mathcal{E}_{III}) &= \mathbb{P}(|\mathcal{S}^{\tau}|/\tau \geq \rho, \nu \leq \tau \leq \nu + n-1) \notag \\
 &\leq \mathbb{P}(|\mathcal{S}^{\nu - 1}| \geq \nu \rho - n - 1). \notag
\end{align}

Given $\nu >\sqrt{A}$, from Markov's inequality we have
\begin{align}
\mathbb{P}(|\mathcal{S}^{\nu - 1}| \geq \nu \rho - n - 1|\nu) \leq \frac{ \mathbb{E}|\mathcal{S}^{\nu - 1}|} {\nu \rho - n - 1 } \label{E222}
\end{align}
where $\mathbb{E}|\mathcal{S}^{\nu- 1}|$ denotes the expected number of samples taken by the decoder before time $\nu$.

In the proposed sampling strategy, the expected number of samples taken by the decoder in each block, denoted as $\mathbb{E}N$, is
\begin{align}\label{EA011}
\mathbb{E}N =\Delta_{1} + \sum_{i=2}^{l}(\prod_{j=1}^{i-1}{p_{j})\Delta_{i}}\leq f(n)^{\delta} + \sum_{i=2}^{l}e^{-(\beta_{i}-c_{i})}
\end{align}
where $p_{j}$ denotes the probability of false-alarm in the $j$th phase, {\it{i.e.}}
$$p_j=\mathbb{P}_{\star}(r(Y^{\Delta_j})>\beta_j).$$
Therefore, $\mathbb{E}|\mathcal{S}^{\nu - 1}|$ can be upper bounded by
\begin{align}
\mathbb{E}|\mathcal{S}^{\nu - 1}| &\leq \frac{\nu - 1}{\Delta(n)}\mathbb{E}N \notag \\
 &= \frac{\nu - 1}{n}f\left(n\right)^{1-\delta}\left(1+f\left(n\right)^{-\delta} \sum_{i=2}^{l}e^{-(\beta_{i}-c_{i})}\right). \label{E221}
\end{align}

Plugging (\ref{E221}) into (\ref{E222}) we get
\begin{align}
\mathbb{P}(|\mathcal{S}^{\nu - 1}| \geq \nu \rho- &n - 1|\nu) \notag   \\
&\leq \frac{ \left(1+f(n)^{-\delta} \sum_{i=2}^{l}e^{-(\beta_{i}-c_{i})}\right) }{f(n)^{\delta}(1-n^{2}/(f(n)\nu))}. \label{E771A}
\end{align}

Therefore,
\begin{align}
&\mathbb{P}(\mathcal{E}_{III}) \notag \\
&\leq \mathbb{P}\left(\nu \leq \sqrt{A}\right) + \mathbb{P}\left(|\mathcal{S}^{\nu-1}| \geq \nu\rho-n-1,\nu > \sqrt{A}\right) \notag \\
&\leq \frac{1}{\sqrt{A}} + \frac{ \left(1+f(n)^{-\delta} \sum_{i=2}^{l}e^{-(\beta_{i}-c_{i})}\right) }{f(n)^{\delta}(1-n^{2}/\sqrt{A})} \label{E771B}
\end{align}
where \eqref{E771B} follows by \eqref{E771A} and the fact that $\nu$ is uniformly distributed over $\{1,2,\cdots,A\}.$ Inequality \eqref{E414} follows.
\section{proof of Inequality (\ref{E708})}\label{AppB}


To simplify the notation, we first define event $\mathcal{G}_j$ as
\begin{align}\label{E709}
\mathcal{G}_j \de \{ i(C^{n-\Delta(n)}(m);Y_{j}^{n-\Delta(n)}) \geq \gamma \}
\end{align}
where $j \in \{1,2,\cdots,n-1\}$ and where 
$$Y_{j}^{n-\Delta(n)} \overset{\text{def}}{=} {Y}_{\nu + \Delta(n)-j}{Y}_{\nu + \Delta(n)-j+1}\cdots{Y}_{\nu + n-1-j}.$$
Note that if event $\mathcal{E}_{IV}$ happens then $\mathcal{G}_j$ occurs for some $j \in \{1,2,\cdots,n-1\}$. 

Let us first consider the case where $1\leq j\leq \Delta(n)$. Since $Y_{j}^{n-\Delta(n)}$ is independent of $C^{n-\Delta(n)}(m)$ by a standard property of information density (see \cite[Eq. $(43)$]{polyanskiy2013asynchronous}) we have
\begin{align}\label{EA112}
\mathbb{P}(\mathcal{G}_j) \leq e^{-\gamma}\qquad 1\leq j\leq \Delta(n).
\end{align}
Now let us consider the case $\Delta(n)+1 \leq j \leq n - 1$. In this case $Y_{j}^{n-\Delta(n)}$ is induced partly by the idle symbol and partly by $C^{n-\Delta(n)}(m)$---therefore $Y_{j}^{n-\Delta(n)}$ depends on $C^{n-\Delta(n)}(m)$.
Following the same arguments as in \cite[Eq. 4.9(b)]{gray1976sliding} it follows that event \eqref{E709} can be upper bounded as
\begin{align}\label{EA111}
\mathbb{P}(\mathcal{G}_j) \leq Me^{-\gamma} \qquad \Delta(n)+1 \leq j \leq n - 1.
\end{align}
Finally from \eqref{EA112} and \eqref{EA111} and a union bound over time indices we get \eqref{E708b}.

\section{Proof of Lemma~\ref{Lem2}\label{AppD}}

\subsection{Achievability}

The achievability part can be established by applying \cite[Theorem $45$]{polyanskiy2010channel} with distribution $P \in \Pi_{\alpha}$ chosen to achieve~$V_{\varepsilon}(\alpha)$.

\subsection{Converse} 

For the converse we extend the argument of \cite[converse of Theorem $48$]{polyanskiy2010channel} to include the constraint that $P \in \Pi_{\alpha}$. To do this we need to show that the following property (required in \cite[Eq.$(502)$]{polyanskiy2010channel}) holds:

\noindent \emph{Property:} For any $ y \in \mathcal{Y}$ and any $P \in \Pi_\alpha$ we have $PW(y) > 0$.\footnote{Recall that in our channel model we assumed that for any $y \in \mathcal{Y}$ there is some $x \in \mathcal{X}$ such that $W(y|x) > 0$.}

To show this we follow the similar arguments as in the proof of \cite[Chap.$4$, Corollary $1$]{gallager1968information} and use the Karush-Kuhn-Tucker (KKT) conditions for characterizing distributions in $ \Pi_{\alpha}$. The optimization problem at hand is:
$$\max. \quad I(P,W) $$
\begin{align}
s.t. \quad \sum_{i = 1}^{|\mathcal{X}|}{P(x_{i})} -1 &= 0\label{E870}\\
P(x_{i}) &\geq 0 \label{E871}  \\
D(PW(\cdot)||W(\cdot|\star))-\alpha &\geq 0. \label{E872}  
\end{align}
The KKT conditions at any point $P^{*} \in \Pi_{\alpha}$ are:
\begin{align}
P^{*}(x_{i})  &\geq 0,  \notag \\
\sum_{i = 1}^{|\mathcal{X}|}{P^{*}(x_{i}) -1} &= 0,\notag \\
\lambda &\geq 0, \notag \\
\lambda_{i} &\geq 0,  \notag \\
\lambda(D(P^{*}W(\cdot)||W(\cdot|\star))-\alpha)&= 0, \notag \\
\lambda_iP^{*}(x_i)&=0, \notag \\
I_{x_{i}}(P^{*},W) - \lambda\Big(\sum_{y \in \mathcal{Y}}W(y|x_i)&\log{\frac{PW(y)}{W(y|\star)}}+1 \Big) \notag \\
+1+\lambda_{i} - \mu&=0, \label{E701} 
\end{align}
where $i \in\{ 1,\cdots,|\mathcal{X}|\}$, where $$I_{x_{i}}(P^{*},W) \overset{\text{def}}{=} D(W(\cdot|x_{i})||P^{*}W(\cdot))$$ and where $\mu$, $\lambda_i$'s, and $\lambda$ are the Lagrange multipliers associated with constraints (\ref{E870}), (\ref{E871}), and (\ref{E872}), respectively. 

Taking the expectation in \eqref{E701} under distribution $P^{*}$ we get 
\begin{align}\label{101}
C(\alpha) - \lambda(\alpha+1)+1-\mu=0.
\end{align}

Substituting \eqref{101} into \eqref{E701} we get 
\begin{align}
I_{x_{i}}(P^{*},W) = C(\alpha) - \lambda\left(\alpha - \sum_{y \in \mathcal{Y}}W(y|x_i)\log{\frac{PW(y)}{W(y|\star)}}\right)\label{E702}
\end{align}
when $P^{*}(x_{i})>0$ and 
\begin{align}
I_{x_{i}}(P^{*},W) \leq C(\alpha) - \lambda\left(\alpha - \sum_{y \in \mathcal{Y}}W(y|x_i)\log{\frac{PW(y)}{W(y|\star)}}\right)\label{E703}
\end{align}
when $P^{*}(x_{i})=0$.

Now suppose by contradiction that there exists some $y \in \mathcal{Y}$ such that $P^*W(y) = 0$. Then for any $x_i \in \mathcal{X}$ satisfying $W(y|x_i) > 0$ we have $I_{x_i}(P^{*},W)=\infty$. However, the right-hand sides of $I_{x_i}(P,W)$ in \eqref{E702} and \eqref{E703}  are upper bounded by $C(\alpha)$. This contradiction implies that the above property holds.
\section{}
\begin{lemma} \label{intersect}
Let $q(n)$ be such that $q(n)=\omega(1)$ and $q(n)=o(\sqrt{A_n})$. If $\rho_n=o(1/q(n))$ then
$${\mathbb{P}}({\cal{S}}^\tau \cap\{\nu,\nu+1,\ldots,\nu+q(n)-1\}\ne \emptyset)=o(1).$$ 
\end{lemma}
The proof of the lemma can be found in the converse of \cite[Theorem $5$]{chandar2015quickly} with $B$
and $d_B$ changed to $n$ and $q(n)$, respectively.



\ifCLASSOPTIONcaptionsoff
  \newpage
\fi

\bibliographystyle{unsrt}
\bibliography{references}
\end{document}